\documentclass[10pt,letter,amsmath,amssymb,aps,prd,nofootinbib,notitlepage,superscriptaddress,twocolumn]{revtex4-2}
\usepackage{lmodern}
\usepackage{graphicx,color}
\usepackage[usenames,dvipsnames]{xcolor}
\graphicspath{{figures/}}
\usepackage{comment}
\usepackage{cancel}

\usepackage[T1]{fontenc}
\usepackage[utf8]{inputenc}
\usepackage{color}
\usepackage{amsmath,amssymb}
\usepackage{esint}
\usepackage{tikz}

\usepackage[normalem]{ulem}

\makeatletter

\definecolor{somegreen}{cmyk}{0,0.49,0.98,0.09}
\definecolor{red}{rgb}{1,0,0}
\definecolor{magenta}{cmyk}{0,1,0,0}
\definecolor{lavender}{cmyk}{0.16,0.67,0,0.57}
\definecolor{darkgreen}{rgb}{0,0.65,0.05}
\definecolor{antiquefuchsia}{rgb}{0.33, 0.1, 0.89}

\newcommand{\ok}{$\Omega_{k0}$}
\newcommand{\lcdm}{$\Lambda$CDM}

\usepackage[unicode=true,
 bookmarks=true,bookmarksnumbered=false,bookmarksopen=false,
 breaklinks=false,pdfborder={0 0 1},backref=false,colorlinks=true]
 {hyperref}

\definecolor{urlblue}{rgb}{0,0,0.9}
\definecolor{linkblue}{rgb}{0,0,.8}
\definecolor{linkgreen}{rgb}{0,0.45,0}\definecolor{linkpurple}{rgb}{0.7,0.0,0.4}
\definecolor{linkorange}{rgb}{0.7,0.1,0.0}
\AtBeginDocument{\hypersetup{
linkcolor=linkblue,
citecolor=linkorange,
urlcolor=urlblue}}\definecolor{urlblue}{rgb}{0,0,0.9}\definecolor{linkblue}{rgb}{0,0,.8}\definecolor{linkgreen}{rgb}{0,0.45,0}\definecolor{linkpurple}{rgb}{0.7,0.0,0.4}\definecolor{linkorange}{rgb}{0.7,0.1,0.0}
\AtBeginDocument{\hypersetup{
linkcolor=linkblue,
citecolor=linkorange,
urlcolor=urlblue}}

\let\jnl@style=\rm
\def\ref@jnl#1{{\jnl@style#1}}

\def\aj{\ref@jnl{AJ}}                   % Astronomical Journal
\def\actaa{\ref@jnl{Acta Astron.}}      % Acta Astronomica
\def\araa{\ref@jnl{ARA\&A}}             % Annual Review of Astron and Astrophys
\def\apj{\ref@jnl{ApJ}}                 % Astrophysical Journal
\def\apjl{\ref@jnl{ApJ}}                % Astrophysical Journal, Letters
\def\apjs{\ref@jnl{ApJS}}               % Astrophysical Journal, Supplement
\def\ao{\ref@jnl{Appl.~Opt.}}           % Applied Optics
\def\apss{\ref@jnl{Ap\&SS}}             % Astrophysics and Space Science
\def\aap{\ref@jnl{A\&A}}                % Astronomy and Astrophysics
\def\aapr{\ref@jnl{A\&A~Rev.}}          % Astronomy and Astrophysics Reviews
\def\aaps{\ref@jnl{A\&AS}}              % Astronomy and Astrophysics, Supplement
\def\azh{\ref@jnl{AZh}}                 % Astronomicheskii Zhurnal
\def\baas{\ref@jnl{BAAS}}               % Bulletin of the AAS
\def\bac{\ref@jnl{Bull. astr. Inst. Czechosl.}}
\def\caa{\ref@jnl{Chinese Astron. Astrophys.}}
\def\cjaa{\ref@jnl{Chinese J. Astron. Astrophys.}}
\def\icarus{\ref@jnl{Icarus}}           % Icarus
\def\jcap{\ref@jnl{J. Cosmology Astropart. Phys.}}
\def\jrasc{\ref@jnl{JRASC}}             % Journal of the RAS of Canada
\def\memras{\ref@jnl{MmRAS}}            % Memoirs of the RAS
\def\mnras{\ref@jnl{MNRAS}}             % Monthly Notices of the RAS
\def\na{\ref@jnl{New A}}                % New Astronomy
\def\nar{\ref@jnl{New A Rev.}}          % New Astronomy Review
\def\pra{\ref@jnl{Phys.~Rev.~A}}        % Physical Review A: General Physics
\def\prb{\ref@jnl{Phys.~Rev.~B}}        % Physical Review B: Solid State
\def\prc{\ref@jnl{Phys.~Rev.~C}}        % Physical Review C
\def\prd{\ref@jnl{Phys.~Rev.~D}}        % Physical Review D
\def\pre{\ref@jnl{Phys.~Rev.~E}}        % Physical Review E
\def\prl{\ref@jnl{Phys.~Rev.~Lett.}}    % Physical Review Letters
\def\pasa{\ref@jnl{PASA}}               % Publications of the Astron. Soc. of Australia
\def\pasp{\ref@jnl{PASP}}               % Publications of the ASP
\def\pasj{\ref@jnl{PASJ}}               % Publications of the ASJ
\def\rmxaa{\ref@jnl{Rev. Mexicana Astron. Astrofis.}}%
\def\qjras{\ref@jnl{QJRAS}}             % Quarterly Journal of the RAS
\def\skytel{\ref@jnl{S\&T}}             % Sky and Telescope
\def\solphys{\ref@jnl{Sol.~Phys.}}      % Solar Physics
\def\sovast{\ref@jnl{Soviet~Ast.}}      % Soviet Astronomy
\def\ssr{\ref@jnl{Space~Sci.~Rev.}}     % Space Science Reviews
\def\zap{\ref@jnl{ZAp}}                 % Zeitschrift fuer Astrophysik
\def\nat{\ref@jnl{Nature}}              % Nature
\def\iaucirc{\ref@jnl{IAU~Circ.}}       % IAU Cirulars
\def\aplett{\ref@jnl{Astrophys.~Lett.}} % Astrophysics Letters
\def\apspr{\ref@jnl{Astrophys.~Space~Phys.~Res.}}
\def\bain{\ref@jnl{Bull.~Astron.~Inst.~Netherlands}}
\def\fcp{\ref@jnl{Fund.~Cosmic~Phys.}}  % Fundamental Cosmic Physics
\def\gca{\ref@jnl{Geochim.~Cosmochim.~Acta}}   % Geochimica Cosmochimica Acta
\def\grl{\ref@jnl{Geophys.~Res.~Lett.}} % Geophysics Research Letters
\def\jcp{\ref@jnl{J.~Chem.~Phys.}}      % Journal of Chemical Physics
\def\jgr{\ref@jnl{J.~Geophys.~Res.}}    % Journal of Geophysics Research
\def\jqsrt{\ref@jnl{J.~Quant.~Spec.~Radiat.~Transf.}}
\def\memsai{\ref@jnl{Mem.~Soc.~Astron.~Italiana}}
\def\nphysa{\ref@jnl{Nucl.~Phys.~A}}   % Nuclear Physics A
\def\physrep{\ref@jnl{Phys.~Rep.}}   % Physics Reports
\def\physscr{\ref@jnl{Phys.~Scr}}   % Physica Scripta
\def\planss{\ref@jnl{Planet.~Space~Sci.}}   % Planetary Space Science
\def\procspie{\ref@jnl{Proc.~SPIE}}   % Proceedings of the SPIE

\makeatother

\makeatother

\begin{document}
\title{Cosmological Spatial Curvature with the Alcock-Paczyński Effect}

\author{Luca Amendola}
\affiliation{Institute of Theoretical Physics, Philosophenweg 16, Heidelberg University, 69120, Heidelberg, Germany}
\author{Marco Marinucci}
\affiliation{Dipartimento di Fisica e Astronomia “G. Galilei”, Università degli Studi di Padova, via Marzolo 8, I-35131, Padova, Italy}
\affiliation{INFN, Sezione di Padova,
Via Marzolo 8, I-35131, Padova, Italy}
\author{Miguel Quartin}
\affiliation{Instituto de Física, Universidade Federal do Rio de Janeiro, 21941-972, Rio de Janeiro, RJ, Brazil}
\affiliation{Observatório do Valongo, Universidade Federal do Rio de Janeiro, 20080-090, Rio de Janeiro, RJ, Brazil}
\affiliation{PPGCosmo, Universidade Federal do Espírito Santo, 29075-910, Vitória, ES, Brazil}
\date{\today }

\begin{abstract}
    We propose a methodology to measure the cosmological spatial curvature by employing the deviation from statistical isotropy due to the Alcock-Paczyński effect of large scale galaxy clustering.  This approach has a higher degree of model independence than most other proposed methods, being independent of calibration of standard candles, rulers, or clocks, of the power spectrum shape (and thus also of the pre-recombination physics), of the galaxy bias, of the theory of gravity,  of the dark energy model and of the background cosmology in general. We find that a combined DESI-Euclid galaxy survey can achieve $\Delta \Omega_{k0}=0.057$ at 1$\sigma$ C.L.~in the redshift range $z<2$ by combining power-spectrum and bispectrum measurements.
\end{abstract}

\maketitle

\textbf{\emph{Introduction}}. The spatial curvature $\Omega_{k0}$ has been one of the most investigated cosmological parameters over the last decades. It is a standard degree of freedom of the Friedman-Lemaître-Robertson-Walker metric, with a very important role in our understanding of the Universe. The possibility of a non-flat universe thus continues to captivate both researchers and laypersons.

The first strong observational constraints on flatness came from the measurements of the first peak of the Cosmic Microwave Background (CMB) angular power spectrum~\cite{Boomerang:2000efg}. Since then, high-resolution maps of the CMB have continued to tighten these constraints, and the current best one comes from the Planck satellite. There is, however, a strong debate on which are the current most reliable measurements. The combination of temperature, polarization and lensing yields $\Omega_{k0}=-0.0106 \pm 0.0065$~\cite{Planck:2018vyg}, consistent with flatness. But the CMB lensing itself is too large to fit the standard $\Lambda$CDM model~\cite{Calabrese:2008rt,Planck:2018vyg}. Dropping lensing, one gets $-0.095 < \Omega_{k0} < -0.007$ at 99\% CL~\cite{DiValentino:2019qzk}, favoring a closed universe (see also~\cite{Handley:2019tkm}). This disagreement highlights the fact that CMB measurements are always performed within a cosmological paradigm the requires assuming a specific model for both the early and late universes.

Curvature can also be measured through its effects on the late-time universe. In particular, it affects measurements of the luminosity ($D_L$) and angular diameter ($D_A$) distances, of the  expansion rate ($H(z)$) and of both weak and strong lensing. Numerous works analyzed combinations of these observables to constrain curvature independently of the CMB. Several of these made use of the so-called cosmic chronometers (CC) to infer $H(z)$ and constrained $\Omega_{k0}$ by combining $H(z)$ with supernova distances (SN)~\cite{Sapone:2014nna,Cai:2015pia,Li:2016wjm,Jesus:2019jvk,Cao:2021ldv,Dhawan:2021mel,Favale:2023lnp}, with the Baryonic Acoustic Oscillation scale (BAO)~\cite{Yu:2017iju,Ryan:2018aif,Cao:2021ldv,Favale:2023lnp}, or with lensing~\cite{Rana:2016gha}. The obtained uncertainties on \ok~are around $\sim0.1-0.2$. However, CC are based on modeling passively evolving galaxies, and their accuracy level is still under debate~\cite{Moresco:2022phi}. Without CC, constraints on \ok~were also obtained with precision $\sim0.5-0.9$ combining supernova distances and lensing~\cite{Rasanen:2014mca}.  Forecasts on $\Omega_{k0}$ have also been performed using weak-lensing from Euclid or the Large Synoptic Survey Telescope~\cite{Leonard:2016evk}; intensity mapping~\cite{Witzemann:2017lhi}; supernovae and BAO~\cite{Dhawan:2021mel};  standard sirens~\cite{Wei:2018cov,Alfradique:2022tox} and the clustering of standard candles~\cite{Quartin:2021dmr,Alfradique:2022tox}.

A promising avenue to avoid CC relies on measurements of the large-scale structure (LSS) alone, since radial and transversal correlations allow measurements of both $D_A(z)$ and $H(z)$. The recent Dark Energy Spectroscopic Instrument (DESI) 2024 results using the BAO alone obtained different estimates according to the assumed model: $\Omega_{k0} = 0.065^{+0.068}_{-0.078}$ ($0.087^{+0.100}_{-0.085}$) assuming the \lcdm~(o$w_0 w_a$CDM) model~\cite{DESI:2024mwx}. More information can be obtained using the full power spectrum $P(k)$ shape.  So called full-shape analyses have proven to be a fruitful source of information about the cosmological parameters~\cite{Ivanov:2019pdj, Kobayashi:2021oud, DAmico:2019fhj, DAmico:2022osl, Ivanov:2023qzb}. Recent Baryon Oscillation Spectroscopic Survey (BOSS) full-shape analysis for $\Lambda$CDM resulted in $\Omega_{k0} = -0.044^{+0.043}_{-0.044}$~\cite{Chudaykin:2020ghx}.

One important recent concern in the field has been to push for model-independent measurements.  BAO measurements need the sound horizon scale $r_d$ to estimate $H,D_A$, so they cannot be considered a model-independent method. The combination $H D_A$ is independent of $r_d$ but it alone cannot constrain $\Omega_{k0}$ unless one fixes the cosmology. In fact, note that the DESI results above change for the two models quoted. Moreover, as shown in the analysis presented in~\cite{eBOSS:2020hur}, BAO and $P(k)$ analyses have usually been carried out fixing the linear power spectrum to a fiducial $\Lambda$CDM cosmology on top of which a specific model template was assumed, adding an extra layer of model dependence.

In this Letter we aim at results that are model-independent both with respect to late-time cosmology (the background expansion) and to early-time cosmology (the power spectrum shape). Non-parametric fits have been employed to mitigate  modeling of late-time cosmology using Gaussian Processes~\cite{Cai:2015pia,Yang:2020bpv,Dhawan:2021mel,Favale:2023lnp}, polynomial fits~\cite{Jesus:2019jvk} or smoothing techniques~\cite{LHuillier:2016mtc}. The option we follow is instead to use directly the measurements of $D_A$ and $H$ in different redshift bins~\cite{Sapone:2014nna,Takada:2015mma}. A late-time model-independence has the advantage of being robust with respect to uncertainties related to dark energy, which is important since there are hints of a tension with late-universe data (e.g.~\cite{Vagnozzi:2020rcz,DESI:2024mwx}). As we will discuss below,  model-independence  can however also be extended to the early universe cosmology, that is to the power spectrum shape. This  also makes results robust against non-standard early universe physics. In fact, the current Hubble tension has sparked interest in more exotic early universe scenarios as a possible explanation~\cite{Kamionkowski:2022pkx}. Extending model-independence to the early universe means we do not have to  assume that $P(k)$ has the $\Lambda$CDM shape, or is parametrized by a restricted set of parameters, for instance the distortion parameters $\alpha_{\parallel},\alpha_{\perp}$, plus the growth rate $f$ and the normalization $\sigma_8$ as in, e.g., \cite{Gil-Marin:2016wya} where, moreover, the non-linear corrections were evaluated only for the $\Lambda$CDM model. To the best of our knowledge, no work so far has investigated the possibility of measuring the spatial curvature in the same model-independent way that we propose in this Letter.

In this work, we employ the FreePower method~\cite{Amendola:2019lvy,Amendola:2022vte,Amendola:2023awr,Schirra:2024rjq}. There are two  main differences between FreePower and other approaches. The first is that  in FreePower we leave the linear power spectrum shape free to vary in several wave bands, rather than adopting a parametrization based on cosmological models. We make use of the Alcock-Paczyński (AP) effect, which depends only on  a combination of the dimensionless expansion rate $E$ and dimensionless comoving angular diameter distance $L_A$
\begin{equation}
    E(z)\equiv H(z)/H_0\,,\quad L_A(z)\equiv H_0 D_A (z)\,.
\end{equation}
We note that one can distinguish between an anisotropic AP effect that depends only on $EL_A$ and an isotropic distortion that depends on $L_A$. The cosmological information is then extracted in a  geometrical way through these effects  which distort the  multipole structure of the spectrum. The second main difference is that instead of choosing a particular cosmological model, we leave free to vary also the functions $f$, $E$ and $L_A$ in each redshift bin, together with all necessary nuisance parameters. The only assumptions of the method are that the background metric is FLRW, and that both power spectrum $P(k)$ and bispectrum can be expanded in a  perturbation expansion that is sufficiently accurate up to some wavelength. We take as forecasted data the one-loop $P(k)$ and tree-level bispectrum of galaxy clustering.

We adopt for the non-linear kernels the  expressions derived in~\cite{DAmico:2021rdb}, which, while relying on perturbativity, are based on general considerations of symmetry rather than on specific models. Our basic parameters are then 25 values of the linear $P(k_i)$ in the $k$ interval $0.01-0.25 \,h/$Mpc, plus, for each redshift bin,  $f$, $E$ and $L_A$,  seven bias and bootstrap parameters, a smoothing velocity dispersion and a counterterm parameter, and finally three shot noises. In total, we have 15 parameters for each redshift bin plus 25 $k$-band parameters. We adopt two cutoff schemes: a more ``aggressive'' one, which is our default scheme, in which we take $k_{\rm max}^{\rm P}=0.25 \, h/$Mpc for the power spectrum and  $k_{\rm max}^{\rm B}=0.1 \,h/$Mpc for the bispectrum; and a ``conservative'' one, in which the two cutoffs are $0.20\,h/$Mpc and $0.08\,h/$Mpc, respectively.

It is crucial to remark that while the combination $E\,L_A$ can be measured via the AP effect also in the linear regime,
the two quantities $E,L_A$ cannot be disentangled since $L_A$ is degenerate with $b_1^2(z) G^2(z)P(k)$, where $b_1$ is the linear bias and $G$ is the linear growth function. The non-linear corrections, however, introduce  additional dependencies  that break this degeneracy. See Appendix A for more details on how this degeneracy is broken.

Another advantage of the FreePower approach  is that it can derive constraints directly on the dimensionless variables $E$ and $L_A$, and thus directly on $\Omega_{k0}$. This is in contrast with using the popular combination CC and SN or standard sirens, which constrains only the quantity \ok$H_0^2$, and thus requires either an extra probe to constrain $H_0$ or an extrapolation of the $H(z)$ data to $z \rightarrow 0$ to break the degeneracy.

We assume that $f$ does not depend on $k$ in the interval here considered. This is not a fundamental limitation, as we have shown in~\cite{Amendola:2022vte}, but it is a safe approximation in many models (e.g., massive neutrinos induce a variation of $f$ with $k$ of less than 1\% in the viable range, see e.g.~\cite{Kiakotou:2007pz}). More details in Ref.~\cite{Amendola:2023awr}.

Our approach addresses therefore  both the issue of improving both accuracy (being more model-independent than other approaches) and precision (employing the information in the one-loop $P(k)$ and in the bispectrum).
%We refer to Ref.~\cite{Amendola:2023awr}  for a detailed description.

\bigskip

\textbf{\emph{Spatial curvature constraints.}}
We applied the FreePower method to produce Fisher matrix (FM) forecasts for a joint DESI and Euclid dataset. The DESI survey \cite{DESI:2016fyo,Vargas-Magana:2018rbb} is a ground telescope which will produce a spectroscopic map covering 14000 deg$^{2}$ of the sky, covering the range $z=(0-1.6)$ with a combination of the surveys of bright galaxy (BGS),  luminous red galaxies (LRG),  and  emission line galaxies (ELG)~\cite{DESI:2024uvr}. The Euclid survey is a space telescope, launched in 2023,  that will map 15000 deg$^{2}$ of the sky~\cite{laureijs2011euclid}, covering the range $z=[0.8-2.0]$. We adopted redshift bins of width $\Delta z=0.2$  centered on the redshifts listed in the tables, and assume negligible cross-bin correlations. We used DESI specifications (only for BGS and low-redshift ELG in order to be conservative) for the bins with $z \le 0.8$ and Euclid for $0.8\le z \le 2.0$. Since we are performing a forecast, we need a fiducial model, that we choose to be  flat $\Lambda$CDM. Our constraints depend on this choice, but the method remains model-independent. When real data will become available, they will replace the fiducial. The main details of the surveys and the fiducial linear bias $b_1$ are displayed in Table \ref{tab:Euclid}. We assume infinite priors on all bias and counterterm parameters following~\cite{Matos:2023jkn}, where all fiducials and remaining priors can be found. The main difference with respect to~\cite{Amendola:2023awr} is the inclusion of the low-$z$ DESI bins.

As mentioned above, the AP effect plus the isotropic distortion  affect the wavenumber $k$ and the cosine angle $\mu$ in a way that depends only on the AP-like parameters $h\equiv E/E_r\,$ and $\,l\equiv L_A/L_{A,r}$, where the subscript $r$ refers to the (arbitrary) reference cosmological value adopted to convert distances and angles into $k$ and $\mu$,  such that  $\mu=\mu_{r}h /\alpha$ and $k=\alpha k_{r}$, where~\cite{Magira:1999bn,Amendola:2004be,Samushia:2010ki}
\begin{equation}
    \alpha \,=\, l^{-1}\sqrt{\mu_{r}^{2}(h^{2} l^2-1)+1}\,.\label{eq:alpha}
\end{equation}

\setlength\tabcolsep{4pt}
\begin{table}[]
    %\footnotesize
    \small
    \centering
    \begin{tabular}{ccccc}
    \hline
    & $z$ & $V$  &  $10^{3}\,n_g$ & $b_1$  \\
    & &\!\![Gpc$/h]^3\!\!$&  $[h/{\rm Mpc}]^{3}$ & \\
    \hline \hline
    & 0.1 & 0.263 & 118. & 1.41  \\
    & 0.3 & 1.53 & 11.9 & 1.57  \\
    \raisebox{0pt}[0pt][0pt]{\rotatebox[origin=c]{90}{\;\;\;\;\,DESI}} & 0.5 & 3.33 & 1.14 & 1.74 \\
    & 0.7 & 5.15 & 1.07 & 1.15  \\
    \hline
    & 0.9 & 7.22 & 1.54 & 1.26  \\
    & 1.1 & 8.61 & 0.891 & 1.34 \\
    & 1.3 & 9.66 & 0.521 & 1.42  \\
    \raisebox{0pt}[0pt][0pt]{\rotatebox[origin=c]{90}{\;\;\;\;\,Euclid}}& 1.5 & 10.4 & 0.274 & 1.5  \\
    & 1.7 & 11. & 0.152 & 1.58  \\
    & 1.9 & 11.3 & 0.0899 & 1.66  \\    \hline
    \end{tabular}
    \caption{
    Our forecast specifications and fiducials, based on DESI and Euclid forecasts for the full surveys. We use DESI BGS for $z<0.6$, DESI ELG for $0.6<z<0.8$ and Euclid ELG for $z>0.8$. For $z<0.6$ we take values for the bias parameters from the low-z BOSS results~\cite{Ivanov:2019pdj}, while for $z>0.6$ we use the models in~\cite{Chudaykin:2019ock}.
    }
    \label{tab:Euclid}
\end{table}

Once we marginalize over all the other parameters, we see that we can measure $h(z),l(z)$ down to 2\%--3\% in several  redshift bins  in the  aggressive case, as we show in Table~\ref{tab:Euclid-results}.  The marginalized Fisher matrix  for $h,l$ is the main input for the next section.

The spatial curvature is related to $E(z)=H(z)/H_0$ and $L_A(z)$ by the relation  (see e.g.~\cite{Clarkson:2007pz})
\begin{equation}
    \Omega_{k0}=\frac{(L_{A,z}E)^{2}-1}{L_A^{2}}\,.
\end{equation}
This relation depends only on the geometrical properties of the FLRW metric and is valid at any redshift regardless of the specific cosmology and theory of gravity.
In our $h,l$ variables this reads
\begin{equation}
    \Omega_{k0}=\frac{\big(h (\partial_z l) E_r L_{A,r}+ hl\big)^{2}-1}{l^{2}L_{A,r}^{2}} \, ,
    \label{omegak}
\end{equation}
where $\partial_z$ means derivative with respect to $z$. Notice that in terms of $E(z)$ and $L_A(z)$, $\Omega_{k0}$ is independent of $H_0$, as we discussed previously. Also, the expression above is valid for both open and closed curvatures. Therefore, once we measure $E,L_A$, we can also measure $\Omega_{k0}$.

\setlength\tabcolsep{3.5pt}
\begin{table}[]
%\footnotesize
\small
\centering
\begin{tabular}{cccc}
    \hline\\[-8pt]
    $z$ & $\Delta f/f$ & $\Delta h/h$ & $ \Delta l/l$   \\[2pt]
    \hline
 \hline
0.1 & 0.072 & 0.038 & 0.024  \\
 0.3 & 0.052 & 0.023 & 0.015  \\
 0.5 & 0.044 & 0.019 & 0.013 \\
 0.7 & 0.034 & 0.022 & 0.018  \\
 0.9 & 0.031 & 0.019 & 0.016  \\
 1.1 & 0.03 & 0.018 & 0.015  \\
 1.3 & 0.032 & 0.019 & 0.015  \\
 1.5 & 0.039 & 0.021 & 0.017  \\
 1.7 & 0.049 & 0.025 & 0.02  \\
 1.9 & 0.063 & 0.031 & 0.026 \\
   \hline
    \end{tabular}
\caption{Fully marginalized relative error forecast for all redshift bins, %marginalized over all the power spectrum wavebands and the higher-order parameters, aggressive case
(see~\cite{Amendola:2023awr} for a full description).
Here and in then rest of this Letter we report 1$\sigma$ uncertainties.}
\label{tab:Euclid-results}
\end{table}

\begin{figure}[t]
    \centering
    \includegraphics[width = .95\columnwidth]{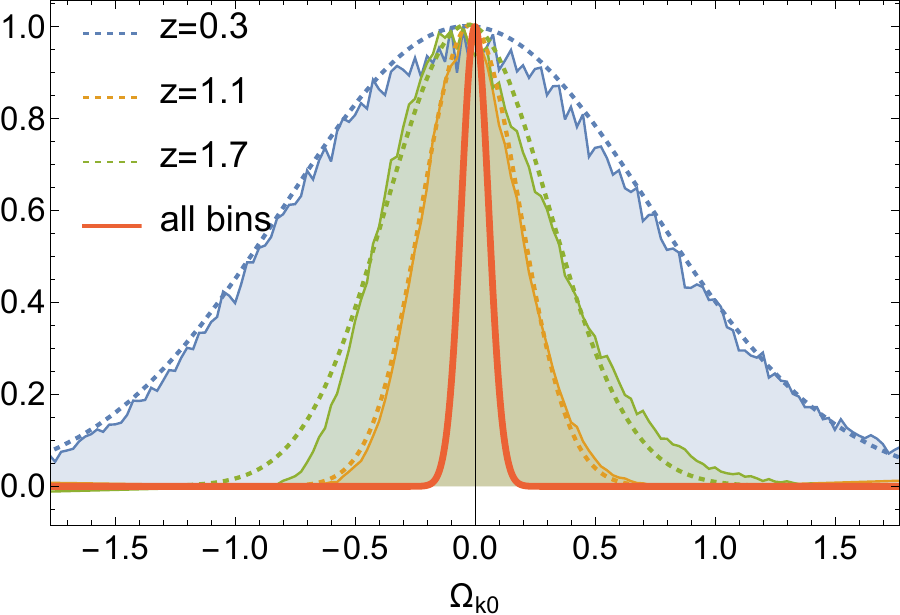}
    \caption{ Numerical distribution of $\Omega_{k0}$ for two redshift bins. The dashed curves are Gaussian fits. The continuous thicker red curve is the likelihood when combining all bins.
    } \label{fig:omkgaussian}
\end{figure}

We need then to propagate  the  constraints on $h,l$ from each bin, including their correlation, to $\Omega_{k0}$. Since $\Omega_{k0}$ depends in a non-linear way on the variables $h,l$, we choose to propagate the errors numerically. We generate $10^5$ random values of $h,l$ in $n_B$ redshift bins $i=1,...,n_B$ from a Gaussian multivariate distribution with means 1 in each bin and covariance given by the inverse of our marginalized FM for $h,l$. Then we discretize Eq.~\eqref{omegak}
\begin{equation}
    \Omega_{k0,i}=\frac{\left[h_i \frac{ (l_{i+1}-l_{i-1})}{2\Delta z} E_r(z_i)L_{A,r}(z_i)+h_il_i\right]^{2}-1}{l_i^{2}L_{A,r}(z_i)^{2}} \,,
    \label{omegakd}
\end{equation}
where $\Delta z$ is the bin size, and from every set of $h_i,l_i$ with $i=2,...,n_B-1$ we produce a value of $\Omega_{k0,i}$. The $n$-th Monte Carlo realization of bin $i$ is denoted as $\Omega_{k0,(i,n)}$. These values are correlated. Then we estimate the $(n_B-2)\times (n_B-2)$ covariance matrix of $\Omega_{k0,(i,j)}$:
\begin{equation}
    C_{\Omega_{k0,(i,j)}}=\big\langle\Omega_{k0,(i,m)}\Omega_{k0,(j,n)}\big\rangle_{m,n}\,.
\end{equation}
The errors on $\Omega_{k0}$ for each bin are in Table~\ref{tab:bin}, while in Fig.~\ref{fig:omkgaussian} we show the distribution for some redshift bins. Since the distribution is well approximated by a  Gaussian,  we can safely interpret the errors in the table  in the usual Gaussian way, i.e.~as 68$\%$ confidence regions. The variance of $\Omega_{k0}$ is obtained by projecting the $(n_B-2)\times (n_B-2)$ Fisher matrix $F=C^{-1}_{\Omega_{k0}}$ onto a single $\Omega_{k0}$. The result is simply
\begin{equation}        \sigma^2_{\Omega_{k0}}=\bigg(\sum_{i,j}F_{ij}\bigg)^{-1}\,.
\end{equation}
Finally, we obtain
\begin{equation}
    \sigma_{\Omega_{k0}}=0.057
\end{equation}
at 68$\%$ for the aggressive specifications, and $\sigma_{\Omega_{k0}}=0.075$ for the conservative ones. If only the power spectrum is employed, then we get 0.094. If one artificially takes the limit of infinite galaxy number density, then  the cosmic-variance limited value of  0.033 can be reached. These results, and the comparison with $\Lambda$CDM, are in Table~\ref{tab:res}. Let us remark that these results are not prior-dominated, that is, the priors for each parameter have been  chosen to be much wider than the final constraints.

\setlength\tabcolsep{3.5pt}
\begin{table}
\small
\centering
\begin{tabular}{c|ccccc}
    \hline
    $z$ & FreePower & FreePower & $\Lambda$CDM\\
    & &  CV limit &  $ + \Omega_{k0}$ \\
    \hline
    0.3 & 0.769 & 0.735 & 0.250 \\
    0.5 & 0.365 & 0.325 & 0.140 \\
    0.7 & 0.244 & 0.207 & 0.110 \\
    0.9 & 0.240 & 0.198 & 0.097 \\
    1.1 & 0.205 & 0.161 & 0.090 \\
    1.3 & 0.218 & 0.143 & 0.089 \\
    1.5 & 0.263 & 0.134 & 0.095 \\
    1.7 & 0.351 & 0.129 & 0.110 \\
    \hline
    \!combined & 0.0572 & 0.0335 & 0.033\\
    \hline
\end{tabular}
\caption{Forecast uncertainties on $\Omega_{k0}$ for each bin (and combined).
We also show the constraints for the standard full-shape approach assuming $\Lambda$CDM.
\label{tab:bin}}
\end{table}

How do these numbers compare with other methods with some degree of model independence? Forecasts for Euclid data using the standard full-shape approach, which assumes a parametrized shape of $P(k)$, and assuming linear theory was valid up to (an optimistic) $k_{\rm max}^{\rm P} = 0.20h$/Mpc were performed in~\cite{Sapone:2014nna}. They found $\sigma$(\ok) around $0.1-0.2$ in 13 redshift bins, which if combined would result in  $\sigma(\Omega_{k0})= 0.033$. A forecast for Euclid + DESI was performed in~\cite{Dhawan:2021mel} using the radial BAO scale combined with Nancy Roman SN, resulting in $\sigma(\Omega_{k0})=0.026$, but this assumes the BAO scale does not evolve. Forecasts for 21cm intensity mapping for HIRAX combining with the CMB distance scale were also performed in~\cite{Witzemann:2017lhi}, resulting in $\sigma(\Omega_{k0})=0.0085$ for an agnostic binned $w(z)$ dark-energy model. Of course, assuming both an early and late-time model allows tighter constraints. For instance, the same HIRAX+CMB constraints shrink to $\sigma(\Omega_{k0})=0.0028$ assuming $w$CDM. Other methods also become very precise. Assuming the $\Lambda$CDM model, using the clustering of Einstein Telescope bright sirens and DESI BGS, $\sigma(\Omega_{k0})=0.018$ was forecast by~\cite{Alfradique:2022tox}, while combining upcoming CMB with Euclid BAO and weak-lensing could yield $\sigma(\Omega_{k0})=0.0018$ (degrading to $0.0088$ for the $w$CDM model)~\cite{Leonard:2016evk}.

We emphasize that, in contrast with our approach, all these constraints have been obtained either assuming specific parametrizations, or the reliability, accuracy and correct calibration in general of standard candles and clocks. For CC in particular, this requires assuming the reliability and robustness of stellar population synthesis models, which form the basis of the method, and that all CC systematic effects can be kept under control.

\begin{table}
    \centering
    %\normalsize
    \small
    \begin{tabular}{lc}
    \hline
    method & combined $\sigma_{\Omega_{k0}}$\tabularnewline
    \hline
    FreePower P+B  & $0.057$ \\
    FreePower conservative P+B & 0.075 \\
    FreePower CV limit P+B & 0.033 \\
    FreePower only P & 0.094 \\
    $\Lambda$CDM full shape P+B & 0.033 \\
    $\Lambda$CDM full shape only P & 0.037 \\
    $\Lambda$CDM full shape P+B+CMB & 0.0021 \\
    \hline
    \end{tabular}
    \caption{Results combining all redshift bins. Note that CMB can only be added by considering a model for both early and late times. P stands for using the power spectrum alone, while P+B  adds also the bispectrum. CV limit denotes the cosmic variance limit, i.e.~$n_g \rightarrow \infty$.\label{tab:res}}
\end{table}

\begin{figure}[t]
    \begin{centering}
    \includegraphics[width= .95\columnwidth]{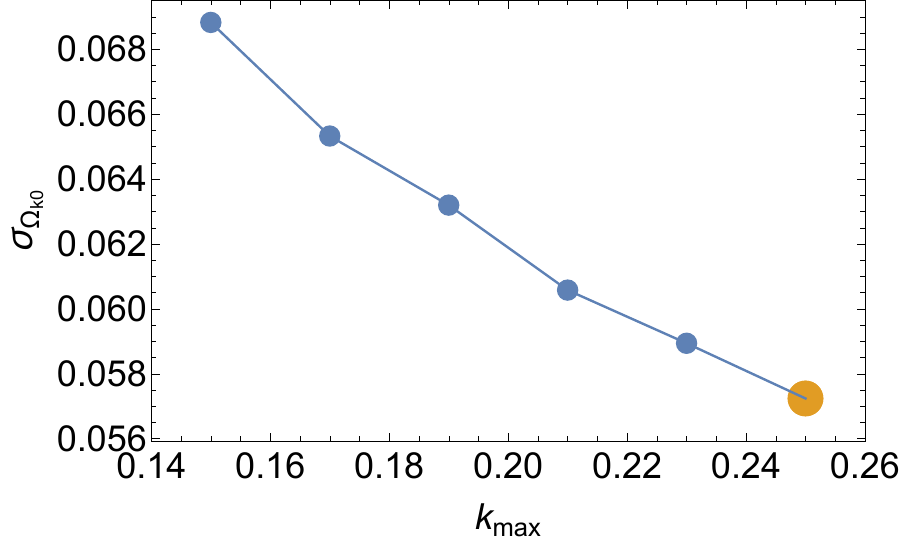}
    \par\end{centering}
    \caption{\label{fig:ok-kmax} Scaling of  $\sigma_{\Omega_{k0}}$ versus $k_{\rm max}^{\rm P}$ for the spectrum (keeping fixed to $k_{\rm max}^{\rm B}=0.10h/$Mpc the bispectrum value). The larger yellow dot represents our reference value. The method has only a weak sensitivity to $k_{\rm max}^{\rm P}$ values.
    }
\end{figure}

Finally, in Fig.~\ref{fig:ok-kmax}, we show how the uncertainty on $\Omega_{k0}$ decreases with an increasing power spectrum cutoff (keeping the bispectrum cutoff at $k_{\rm max}^{\rm B} 0.1\,h/$Mpc).

\begin{figure*}[t]
    \centering
    \includegraphics[width=.72\columnwidth]{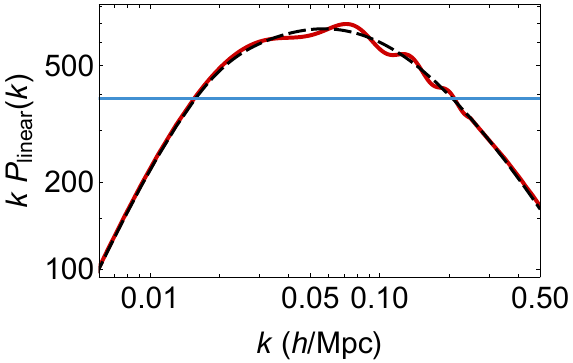}\qquad
    \includegraphics[width=.72\columnwidth]{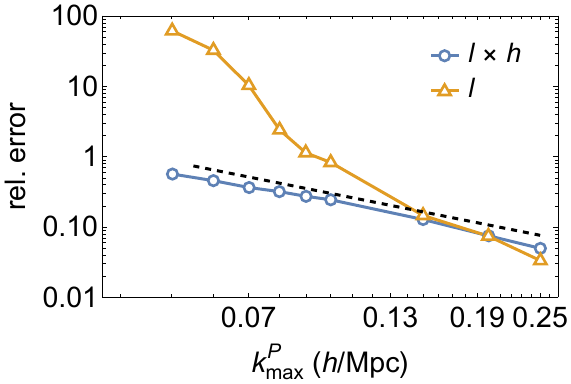}\\
    \includegraphics[width=.72\columnwidth]{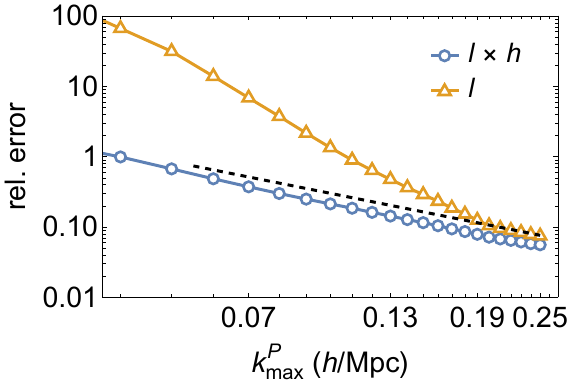}\qquad
    \includegraphics[width=.72\columnwidth]{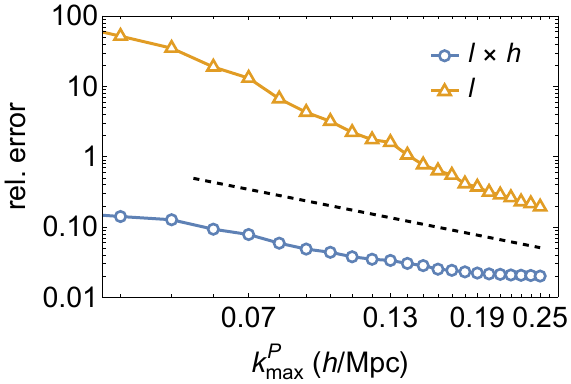}
    \caption{
    Constraints in $h\times l$ and $l$ ($y$-axis) for different values of $k_{\rm max}^{\rm P}$  \emph{Top left:} the three linear $P(k)$ fiducials.
    \emph{Top right:} The standard $\Lambda$CDM case, used in the main text.
    \emph{Bottom left:} case of $P(k)$ without wiggles.
    \emph{Bottom right:} case of $P(k) \propto k^{-1}$.
    The absence of wiggles makes the precision worse (especially for low $k_{\max}$), but the degeneracy is broken nevertheless. For the pure power-law case, a large degeneracy remains ($\sigma(l) \gg \sigma(h\times l)$) since the only scale involved is that of the non-linearity. In all cases $h\times l$ follows the $k_{\rm max}^{-3/2}$ scaling (dashed line), as expected from the simple counting of available modes.
    \label{fig:eta-l} }
\end{figure*}

\bigskip

\textbf{\emph{Discussion}.} We presented a methodology to measure the late-time cosmic spatial curvature that is independent of calibration of standard candles, clocks, or rulers, of the cosmological background, and of models of the power spectrum shape and growth. This model-independent approach makes use of the statistical isotropy of the Universe embedded in the linear and non-linear power spectrum and bispectrum of galaxy clustering. We find that a combination of the DESI and Euclid surveys can constrain $\Omega_{k0}$ to within 0.057, a level competitive with several other less model-independent methods. If $\Omega_{k0}$ is measured   separately in each redshift bin, our approach can also be employed as a test of the Copernican principle \cite{Clarkson:2007pz}.

One can further improve these constraints in a number of ways, e.g.~by adding other redshift bins or larger sky areas, or combining different tracers of structure.

One can also consider external constraints on $H,D$ from standard candles or cosmic chronometers. We tested adding strong external priors for either distance or expansion constraints. We find that FreePower benefits the most from the former: external distance data could improve precision by a factor of almost three. However, this comes  at the cost of assuming these independently measured distances are free of biases and systematic effects in general.

%\section*{Acknowledgements}
\bigskip

\textbf{\emph{Acknowledgments}}. We thank Massimo Pietroni for several discussions and Tjark Harder for help with debugging the numerical code. LA acknowledges support from DFG project  456622116. MM acknowledges support by the MIUR PRIN Bando 2022 - grant 20228RMX4A. MQ is supported by the Brazilian research agencies FAPERJ, CNPq. Many integrals of the FreePower method have been performed using the CUBA routines \cite{Hahn_2005} by T. Hahn (\url{http://feynarts.de/cuba}).

%\appendix*

\bigskip

\textbf{\emph{Appendix A: Impact of the fiducial $P(k)$ and $\Omega_{k0}$}}

The AP effect in the linear regime can only measure the quantity $E L_A$. That is exactly why most works so far have been unable to use the AP effect to measure \emph{both} $h$ and $d$ separately, as we instead do in our method, due to the non-linear effects.  In other words: (i) the isotropic dilatation is degenerate  with the power spectrum shape at the linear level; (ii) therefore one cannot measure both $h$ and $d$; (iii) as an immediate consequence, one cannot measure the spatial curvature, \emph{unless something else is fixed within the context of a model}, e.g. the power spectrum shape and the background cosmology; (iv) in our method, the degeneracy is lifted because of the non-linear correction, so we do not need to fix or parametrize neither the linear spectrum shape nor the background cosmology. This is the crucial advantage of our FreePower approach, and the main novelty of this Letter.

In this Appendix, we investigate the ability of the FreePower method to break this degeneracy. We perform the analysis for three cases: the standard $\Lambda$CDM linear $P(k)$, a case without wiggles and an exotic, pure power-law $P(k)$, chosen as $P(k)\propto k^{-1}$ normalized to have the same $\sigma_8$. In  Fig.~\ref{fig:eta-l} we show the three different $P(k)$ discussed above and the respective errors in $l$ and the combination $hl$. Indeed, the errors depend on the fiducial $P(k)$, but the degeneracy between $l$ and $h$ is broken completely with or without wiggles for high values of $k_{\rm max}^{\rm P}$, when non-linearities kick in. In fact, $k_{\rm max}^{\rm P} \simeq 0.15h$/Mpc is enough to break most of this degeneracy.

In the limiting case of an exact linear power law spectrum, $l$ instead can not be well measured. Nevertheless, it is not completely degenerate, due to the presence of the scale of non-linearity. Similar qualitative results for the combination $h\,l$ have been discussed in~\cite{Amendola:2022vte}.

\begin{figure}[t]
    \centering
    \includegraphics[width = .95\columnwidth]{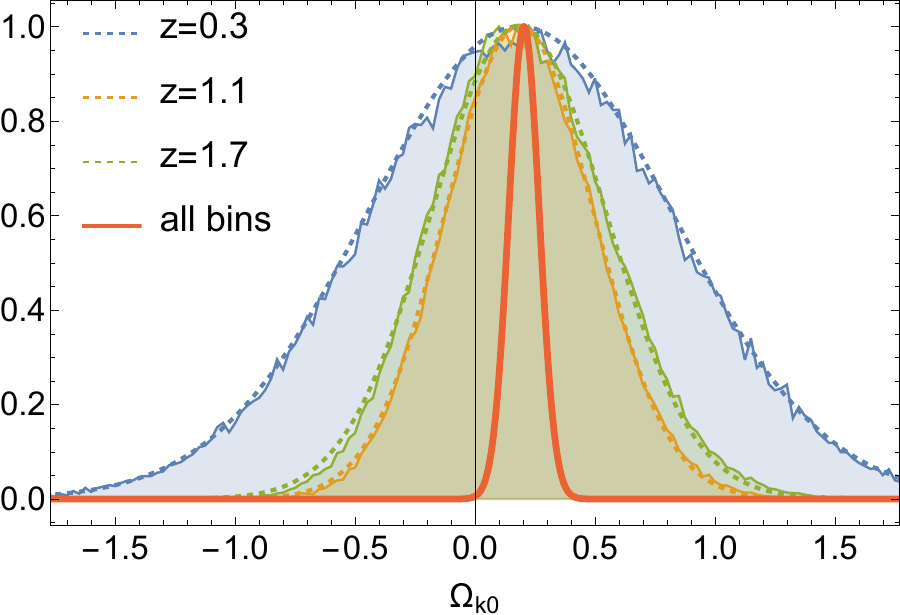}
    \caption{Numerical distribution of $\Omega_{k0}$ for two redshift bins, adopting DESI fiducial  $w_0=-0.827,w_a=-0.75$ and $\Omega_{k0}=0.2$. The dashed curves are Gaussian fits. The continuous thicker red curve is the likelihood when combining all bins.
    } \label{fig:omkDESI}
\end{figure}

As a further proof that the proposed method to measure $\Omega_{k0}$ is independent of the background cosmology, we also adopted a fiducial for $w_0,w_a$ according to the recent DESI results~\cite{DESI:2024mwx}, with spatial curvature $0.2$. Then we generate many values of $E,L_A$ Gaussianly distributed around  this fiducial (in the same bins as our combined DESI+Euclid survey) with relative error $1\%$ and estimate $\Omega_{k0}$ according to Eq.~\eqref{omegakd}. The result given in Fig.~\ref{fig:omkDESI} confirms our method.

\bigskip

\textbf{\emph{Appendix B: $\Omega_{k0}$ constraints assuming $\Lambda$CDM}}
The results shown in this letter are obtained using the FreePower method where the linear power spectrum is not fixed by cosmology, while standard LSS analyses usually assume a specific model, $\Lambda$CDM, \cite{Ivanov:2019pdj, Philcox:2020vvt, Ivanov:2021kcd, Philcox:2022frc, Ivanov:2023qzb, DAmico:2019fhj, DAmico:2022osl}, or its generalizations, e.g.~see~\cite{DAmico:2020tty, Ivanov:2020ril,Piga:2022mge, Taule:2024bot}. We perform a Fisher analysis using the same specifications and fiducial values listed in Table~\ref{tab:Euclid} adopting $\Lambda$CDM plus a non-zero $\Omega_{k0}$. We use the code  PyBird~\cite{DAmico_pybird}, re-adapted to match the biasing scheme of FreePower. We vary simultaneously the cosmology, the bias and the small scales parameters using PyBird (\url{https://github.com/pierrexyz/pybird}).

It has been observed that, at the scales considered, the one-loop power spectrum and the tree-level bispectrum are not very sensitive to some of the Effective Field Theory (EFT) parameters, such as the non-linear bias(es), the counterterms and the shot noise, that are usually fixed or marginalized, see~\cite{Ivanov:2019pdj,DAmico:2019fhj}. For this reason we fix the third order bias, the next-to-next-to-leading order counterterm and the scale dependent shot noise~\cite{DAmico_pybird}, while we leave free to vary all the other bias parameters (a total of seven) in each redshift bin. We consider the monopole and quadrupole of the two statistics, neglecting the correlation among different redshift bins, as this effect has been shown to be negligible~\cite{Bailoni:2016ezz}.

The results of this $\Lambda$CDM forecast are reported in Tables~\ref{tab:bin}--\ref{tab:res}. Adding the bispectrum does not significantly improve the constraints on the curvature parameter: with $P$ alone we obtain $\sigma_{\Omega_{k0}} = 0.037 \;(0.073)$ at 68$\%$ (95$\%$) CL while with $P+B$ we have $\sigma_{\Omega_{k0}} = 0.033 \;(0.065)$ at 68$\%$ (95$\%$) CL. These results are in line with the analysis performed in~\cite{Chudaykin:2020ghx}, and analogous analyses have shown that the bispectrum is essential to constrain the non-linear biases but adds little information about cosmology compared to the $P$ only case~\cite{Philcox:2022frc}. Furthermore, if we include Planck CMB data~\cite{Planck:2018vyg} we obtain $\sigma_{\Omega_{k0}} = 0.0021$, ten times better than LSS alone. This illustrates how, assuming a particular model, the combination of LSS and CMB data is capable of breaking important degeneracies among the cosmological parameters.

\bibliographystyle{JHEP2015}
%\bibliographystyle{mn2e_eprint}
%\bibliography{references,scaling_bib,curvature,references3}
\bibliography{references,curvature,scaling_bib}

\providecommand{\href}[2]{#2}\begingroup\raggedright\begin{thebibliography}{10}

\bibitem{Boomerang:2000efg}
{\bf Boomerang} Collaboration, P.~de~Bernardis et~al., {\it {A Flat universe
  from high resolution maps of the cosmic microwave background radiation}},
  {\em Nature} {\bf 404} (2000) 955--959,
  [\href{http://arxiv.org/abs/astro-ph/0004404}{{\tt astro-ph/0004404}}].

\bibitem{Planck:2018vyg}
{\bf Planck} Collaboration, N.~Aghanim et~al., {\it {Planck 2018 results. VI.
  Cosmological parameters}},  {\em Astron. Astrophys.} {\bf 641} (2020) A6,
  [\href{http://arxiv.org/abs/1807.06209}{{\tt arXiv:1807.06209}}]. [Erratum:
  Astron.Astrophys. 652, C4 (2021)].

\bibitem{Calabrese:2008rt}
E.~Calabrese, A.~Slosar, A.~Melchiorri, G.~F. Smoot, and O.~Zahn, {\it {Cosmic
  Microwave Weak lensing data as a test for the dark universe}},  {\em Phys.
  Rev. D} {\bf 77} (2008) 123531, [\href{http://arxiv.org/abs/0803.2309}{{\tt
  arXiv:0803.2309}}].

\bibitem{DiValentino:2019qzk}
E.~Di~Valentino, A.~Melchiorri, and J.~Silk, {\it {Planck evidence for a closed
  Universe and a possible crisis for cosmology}},  {\em Nature Astron.} {\bf 4}
  (2019), no.~2 196--203, [\href{http://arxiv.org/abs/1911.02087}{{\tt
  arXiv:1911.02087}}].

\bibitem{Handley:2019tkm}
W.~Handley, {\it {Curvature tension: evidence for a closed universe}},  {\em
  Phys. Rev. D} {\bf 103} (2021), no.~4 L041301,
  [\href{http://arxiv.org/abs/1908.09139}{{\tt arXiv:1908.09139}}].

\bibitem{Sapone:2014nna}
D.~Sapone, E.~Majerotto, and S.~Nesseris, {\it {Curvature versus distances:
  Testing the FLRW cosmology}},  {\em Phys. Rev. D} {\bf 90} (2014), no.~2
  023012, [\href{http://arxiv.org/abs/1402.2236}{{\tt arXiv:1402.2236}}].

\bibitem{Cai:2015pia}
R.-G. Cai, Z.-K. Guo, and T.~Yang, {\it {Null test of the cosmic curvature
  using $H(z)$ and supernovae data}},  {\em Phys. Rev. D} {\bf 93} (2016),
  no.~4 043517, [\href{http://arxiv.org/abs/1509.06283}{{\tt
  arXiv:1509.06283}}].

\bibitem{Li:2016wjm}
Z.~Li, G.-J. Wang, K.~Liao, and Z.-H. Zhu, {\it {Model-independent estimations
  for the curvature from standard candles and clocks}},  {\em Astrophys. J.}
  {\bf 833} (2016), no.~2 240, [\href{http://arxiv.org/abs/1611.00359}{{\tt
  arXiv:1611.00359}}].

\bibitem{Jesus:2019jvk}
J.~F. Jesus, R.~Valentim, P.~H. R.~S. Moraes, and M.~Malheiro, {\it {Kinematic
  Constraints on Spatial Curvature from Supernovae Ia and Cosmic
  Chronometers}},  {\em Mon. Not. Roy. Astron. Soc.} {\bf 500} (2020), no.~2
  2227--2235, [\href{http://arxiv.org/abs/1907.01033}{{\tt arXiv:1907.01033}}].

\bibitem{Cao:2021ldv}
S.~Cao, J.~Ryan, and B.~Ratra, {\it {Using Pantheon and DES supernova, baryon
  acoustic oscillation, and Hubble parameter data to constrain the Hubble
  constant, dark energy dynamics, and spatial curvature}},  {\em Mon. Not. Roy.
  Astron. Soc.} {\bf 504} (2021), no.~1 300--310,
  [\href{http://arxiv.org/abs/2101.08817}{{\tt arXiv:2101.08817}}].

\bibitem{Dhawan:2021mel}
S.~Dhawan, J.~Alsing, and S.~Vagnozzi, {\it {Non-parametric spatial curvature
  inference using late-Universe cosmological probes}},  {\em Mon. Not. Roy.
  Astron. Soc.} {\bf 506} (2021), no.~1 L1--L5,
  [\href{http://arxiv.org/abs/2104.02485}{{\tt arXiv:2104.02485}}].

\bibitem{Favale:2023lnp}
A.~Favale, A.~G\'omez-Valent, and M.~Migliaccio, {\it {Cosmic chronometers to
  calibrate the ladders and measure the curvature of the Universe. A
  model-independent study}},  {\em Mon. Not. Roy. Astron. Soc.} {\bf 523}
  (2023), no.~3 3406--3422, [\href{http://arxiv.org/abs/2301.09591}{{\tt
  arXiv:2301.09591}}].

\bibitem{Yu:2017iju}
H.~Yu, B.~Ratra, and F.-Y. Wang, {\it {Hubble Parameter and Baryon Acoustic
  Oscillation Measurement Constraints on the Hubble Constant, the Deviation
  from the Spatially Flat \ensuremath{\Lambda}CDM Model, the
  Deceleration\textendash{}Acceleration Transition Redshift, and Spatial
  Curvature}},  {\em Astrophys. J.} {\bf 856} (2018), no.~1 3,
  [\href{http://arxiv.org/abs/1711.03437}{{\tt arXiv:1711.03437}}].

\bibitem{Ryan:2018aif}
J.~Ryan, S.~Doshi, and B.~Ratra, {\it {Constraints on dark energy dynamics and
  spatial curvature from Hubble parameter and baryon acoustic oscillation
  data}},  {\em Mon. Not. Roy. Astron. Soc.} {\bf 480} (2018), no.~1 759--767,
  [\href{http://arxiv.org/abs/1805.06408}{{\tt arXiv:1805.06408}}].

\bibitem{Rana:2016gha}
A.~Rana, D.~Jain, S.~Mahajan, and A.~Mukherjee, {\it {Constraining cosmic
  curvature by using age of galaxies and gravitational lenses}},  {\em JCAP}
  {\bf 03} (2017) 028, [\href{http://arxiv.org/abs/1611.07196}{{\tt
  arXiv:1611.07196}}].

\bibitem{Moresco:2022phi}
M.~Moresco et~al., {\it {Unveiling the Universe with emerging cosmological
  probes}},  {\em Living Rev. Rel.} {\bf 25} (2022), no.~1 6,
  [\href{http://arxiv.org/abs/2201.07241}{{\tt arXiv:2201.07241}}].

\bibitem{Rasanen:2014mca}
S.~Räsänen, K.~Bolejko, and A.~Finoguenov, {\it {New Test of the
  Friedmann-Lemaître-Robertson-Walker Metric Using the Distance Sum Rule}},
  {\em Phys. Rev. Lett.} {\bf 115} (2015), no.~10 101301,
  [\href{http://arxiv.org/abs/1412.4976}{{\tt arXiv:1412.4976}}].

\bibitem{Leonard:2016evk}
C.~D. Leonard, P.~Bull, and R.~Allison, {\it {Spatial curvature endgame:
  Reaching the limit of curvature determination}},  {\em Phys. Rev. D} {\bf 94}
  (2016), no.~2 023502, [\href{http://arxiv.org/abs/1604.01410}{{\tt
  arXiv:1604.01410}}].

\bibitem{Witzemann:2017lhi}
A.~Witzemann, P.~Bull, C.~Clarkson, M.~G. Santos, M.~Spinelli, and A.~Weltman,
  {\it {Model-independent curvature determination with 21 cm intensity mapping
  experiments}},  {\em Mon. Not. Roy. Astron. Soc.} {\bf 477} (2018), no.~1
  L122--L127, [\href{http://arxiv.org/abs/1711.02179}{{\tt arXiv:1711.02179}}].

\bibitem{Wei:2018cov}
J.-J. Wei, {\it {Model-independent Curvature Determination from
  Gravitational-Wave Standard Sirens and Cosmic Chronometers}},  {\em
  Astrophys. J.} {\bf 868} (2018), no.~1 29,
  [\href{http://arxiv.org/abs/1806.09781}{{\tt arXiv:1806.09781}}].

\bibitem{Alfradique:2022tox}
V.~Alfradique, M.~Quartin, L.~Amendola, T.~Castro, and A.~Toubiana, {\it {The
  lure of sirens: joint distance and velocity measurements with
  third-generation detectors}},  {\em Mon. Not. Roy. Astron. Soc.} {\bf 517}
  (2022), no.~4 5449--5462, [\href{http://arxiv.org/abs/2205.14034}{{\tt
  arXiv:2205.14034}}].

\bibitem{Quartin:2021dmr}
M.~Quartin, L.~Amendola, and B.~Moraes, {\it {The 6~\texttimes{}~2pt method:
  supernova velocities meet multiple tracers}},  {\em Mon. Not. Roy. Astron.
  Soc.} {\bf 512} (2022), no.~2 2841--2853,
  [\href{http://arxiv.org/abs/2111.05185}{{\tt arXiv:2111.05185}}].

\bibitem{DESI:2024mwx}
{\bf DESI} Collaboration, A.~G. Adame et~al., {\it {DESI 2024 VI: Cosmological
  Constraints from the Measurements of Baryon Acoustic Oscillations}},
  \href{http://arxiv.org/abs/2404.03002}{{\tt arXiv:2404.03002}}.

\bibitem{Ivanov:2019pdj}
M.~M. Ivanov, M.~Simonovi\'c, and M.~Zaldarriaga, {\it {Cosmological Parameters
  from the BOSS Galaxy Power Spectrum}},  {\em JCAP} {\bf 05} (2020) 042,
  [\href{http://arxiv.org/abs/1909.05277}{{\tt arXiv:1909.05277}}].

\bibitem{Kobayashi:2021oud}
Y.~Kobayashi, T.~Nishimichi, M.~Takada, and H.~Miyatake, {\it {Full-shape
  cosmology analysis of the SDSS-III BOSS galaxy power spectrum using an
  emulator-based halo model: A 5\% determination of \ensuremath{\sigma}8}},
  {\em Phys. Rev. D} {\bf 105} (2022), no.~8 083517,
  [\href{http://arxiv.org/abs/2110.06969}{{\tt arXiv:2110.06969}}].

\bibitem{DAmico:2019fhj}
G.~D'Amico, J.~Gleyzes, N.~Kokron, K.~Markovic, L.~Senatore, P.~Zhang,
  F.~Beutler, and H.~Gil-Mar\'\i{}n, {\it {The Cosmological Analysis of the
  SDSS/BOSS data from the Effective Field Theory of Large-Scale Structure}},
  {\em JCAP} {\bf 05} (2020) 005, [\href{http://arxiv.org/abs/1909.05271}{{\tt
  arXiv:1909.05271}}].

\bibitem{DAmico:2022osl}
G.~D'Amico, Y.~Donath, M.~Lewandowski, L.~Senatore, and P.~Zhang, {\it {The
  BOSS bispectrum analysis at one loop from the Effective Field Theory of
  Large-Scale Structure}},  \href{http://arxiv.org/abs/2206.08327}{{\tt
  arXiv:2206.08327}}.

\bibitem{Ivanov:2023qzb}
M.~M. Ivanov, O.~H.~E. Philcox, G.~Cabass, T.~Nishimichi, M.~Simonovi\'c, and
  M.~Zaldarriaga, {\it {Cosmology with the galaxy bispectrum multipoles:
  Optimal estimation and application to BOSS data}},  {\em Phys. Rev. D} {\bf
  107} (2023), no.~8 083515, [\href{http://arxiv.org/abs/2302.04414}{{\tt
  arXiv:2302.04414}}].

\bibitem{Chudaykin:2020ghx}
A.~Chudaykin, K.~Dolgikh, and M.~M. Ivanov, {\it {Constraints on the curvature
  of the Universe and dynamical dark energy from the Full-shape and BAO data}},
   {\em Phys. Rev. D} {\bf 103} (2021), no.~2 023507,
  [\href{http://arxiv.org/abs/2009.10106}{{\tt arXiv:2009.10106}}].

\bibitem{eBOSS:2020hur}
{\bf eBOSS} Collaboration, H.~Gil-Marin et~al., {\it {The Completed SDSS-IV
  extended Baryon Oscillation Spectroscopic Survey: measurement of the BAO and
  growth rate of structure of the luminous red galaxy sample from the
  anisotropic power spectrum between redshifts 0.6 and 1.0}},  {\em Mon. Not.
  Roy. Astron. Soc.} {\bf 498} (2020), no.~2 2492--2531,
  [\href{http://arxiv.org/abs/2007.08994}{{\tt arXiv:2007.08994}}].

\bibitem{Yang:2020bpv}
Y.~Yang and Y.~Gong, {\it {Measurement on the cosmic curvature using the
  Gaussian process method}},  {\em Mon. Not. Roy. Astron. Soc.} {\bf 504}
  (2021), no.~2 3092--3097, [\href{http://arxiv.org/abs/2007.05714}{{\tt
  arXiv:2007.05714}}].

\bibitem{LHuillier:2016mtc}
B.~L'Huillier and A.~Shafieloo, {\it {Model-independent test of the FLRW
  metric, the flatness of the Universe, and non-local measurement of
  $H_0r_\mathrm{d}$}},  {\em JCAP} {\bf 01} (2017) 015,
  [\href{http://arxiv.org/abs/1606.06832}{{\tt arXiv:1606.06832}}].

\bibitem{Takada:2015mma}
M.~Takada and O.~Dore, {\it {Geometrical Constraint on Curvature with BAO
  experiments}},  {\em Phys. Rev. D} {\bf 92} (2015), no.~12 123518,
  [\href{http://arxiv.org/abs/1508.02469}{{\tt arXiv:1508.02469}}].

\bibitem{Vagnozzi:2020rcz}
S.~Vagnozzi, E.~Di~Valentino, S.~Gariazzo, A.~Melchiorri, O.~Mena, and J.~Silk,
  {\it {The galaxy power spectrum take on spatial curvature and cosmic
  concordance}},  {\em Phys. Dark Univ.} {\bf 33} (2021) 100851,
  [\href{http://arxiv.org/abs/2010.02230}{{\tt arXiv:2010.02230}}].

\bibitem{Kamionkowski:2022pkx}
M.~Kamionkowski and A.~G. Riess, {\it {The Hubble Tension and Early Dark
  Energy}},  {\em Ann. Rev. Nucl. Part. Sci.} {\bf 73} (2023) 153--180,
  [\href{http://arxiv.org/abs/2211.04492}{{\tt arXiv:2211.04492}}].

\bibitem{Gil-Marin:2016wya}
H.~Gil-Mar\'\i{}n, W.~J. Percival, L.~Verde, J.~R. Brownstein, C.-H. Chuang,
  F.-S. Kitaura, S.~A. Rodr\'\i{}guez-Torres, and M.~D. Olmstead, {\it {The
  clustering of galaxies in the SDSS-III Baryon Oscillation Spectroscopic
  Survey: RSD measurement from the power spectrum and bispectrum of the DR12
  BOSS galaxies}},  {\em Mon. Not. Roy. Astron. Soc.} {\bf 465} (2017), no.~2
  1757--1788, [\href{http://arxiv.org/abs/1606.00439}{{\tt arXiv:1606.00439}}].

\bibitem{Amendola:2019lvy}
L.~Amendola and M.~Quartin, {\it {Measuring the Hubble function with standard
  candle clustering}},  {\em Mon. Not. Roy. Astron. Soc.} {\bf 504} (2021),
  no.~3 3884--3889, [\href{http://arxiv.org/abs/1912.10255}{{\tt
  arXiv:1912.10255}}].

\bibitem{Amendola:2022vte}
L.~Amendola, M.~Pietroni, and M.~Quartin, {\it {Fisher matrix for the one-loop
  galaxy power spectrum: measuring expansion and growth rates without assuming
  a cosmological model}},  {\em JCAP} {\bf 11} (2022) 023,
  [\href{http://arxiv.org/abs/2205.00569}{{\tt arXiv:2205.00569}}].

\bibitem{Amendola:2023awr}
L.~Amendola, M.~Marinucci, M.~Pietroni, and M.~Quartin, {\it {Improving
  precision and accuracy in cosmology with model-independent spectrum and
  bispectrum}},  {\em JCAP} {\bf 01} (2024) 001,
  [\href{http://arxiv.org/abs/2307.02117}{{\tt arXiv:2307.02117}}].

\bibitem{Schirra:2024rjq}
A.~P. Schirra, M.~Quartin, and L.~Amendola, {\it {A model-independent
  measurement of the expansion and growth rates from BOSS using the FreePower
  method}},  \href{http://arxiv.org/abs/2406.15347}{{\tt arXiv:2406.15347}}.

\bibitem{DAmico:2021rdb}
G.~D'Amico, M.~Marinucci, M.~Pietroni, and F.~Vernizzi, {\it {The large scale
  structure bootstrap: perturbation theory and bias expansion from
  symmetries}},  {\em JCAP} {\bf 10} (2021) 069,
  [\href{http://arxiv.org/abs/2109.09573}{{\tt arXiv:2109.09573}}].

\bibitem{Kiakotou:2007pz}
A.~Kiakotou, {\O}.~Elgar{{\o}}y, and O.~Lahav, {\it {Neutrino Mass, Dark
  Energy, and the Linear Growth Factor}},  {\em Phys. Rev. D} {\bf 77} (2008)
  063005, [\href{http://arxiv.org/abs/0709.0253}{{\tt arXiv:0709.0253}}].

\bibitem{DESI:2016fyo}
{\bf DESI} Collaboration, A.~Aghamousa et~al., {\it {The DESI Experiment Part
  I: Science,Targeting, and Survey Design}},
  \href{http://arxiv.org/abs/1611.00036}{{\tt arXiv:1611.00036}}.

\bibitem{Vargas-Magana:2018rbb}
{\bf DESI} Collaboration, M.~Vargas-Maga\~na, D.~D. Brooks, M.~M. Levi, and
  G.~G. Tarle, {\it {Unraveling the Universe with DESI}},  in {\em {53rd
  Rencontres de Moriond on Cosmology}}, pp.~11--18, 2018.
\newblock \href{http://arxiv.org/abs/1901.01581}{{\tt arXiv:1901.01581}}.

\bibitem{DESI:2024uvr}
{\bf DESI} Collaboration, A.~G. Adame et~al., {\it {DESI 2024 III: Baryon
  Acoustic Oscillations from Galaxies and Quasars}},
  \href{http://arxiv.org/abs/2404.03000}{{\tt arXiv:2404.03000}}.

\bibitem{laureijs2011euclid}
{\bf EUCLID} Collaboration, R.~Laureijs, J.~Amiaux, S.~Arduini, J.~L.
  Auguères, J.~Brinchmann, R.~Cole, M.~Cropper, C.~Dabin, et~al., {\it Euclid
  definition study report},  \href{http://arxiv.org/abs/1110.3193}{{\tt
  arXiv:1110.3193}}.

\bibitem{Matos:2023jkn}
I.~Matos, M.~Quartin, L.~Amendola, M.~Kunz, and R.~Sturani, {\it {A
  model-independent tripartite test of cosmic distance relations}},  {\em JCAP}
  {\bf 08} (2024) 007, [\href{http://arxiv.org/abs/2311.17176}{{\tt
  arXiv:2311.17176}}].

\bibitem{Magira:1999bn}
H.~Magira, Y.~P. Jing, and Y.~Suto, {\it {Cosmological redshift-space
  distortion on clustering of high-redshift objects: correction for nonlinear
  effects in power spectrum and tests with n-body simulations}},  {\em
  Astrophys. J.} {\bf 528} (2000) 30,
  [\href{http://arxiv.org/abs/astro-ph/9907438}{{\tt astro-ph/9907438}}].

\bibitem{Amendola:2004be}
L.~Amendola, C.~Quercellini, and E.~Giallongo, {\it {Constraints on perfect
  fluid and scalar field dark energy models from future redshift surveys}},
  {\em Mon. Not. Roy. Astron. Soc.} {\bf 357} (2005) 429--439,
  [\href{http://arxiv.org/abs/astro-ph/0404599}{{\tt astro-ph/0404599}}].

\bibitem{Samushia:2010ki}
L.~Samushia et~al., {\it {Effects of cosmological model assumptions on galaxy
  redshift survey measurements}},  {\em Mon. Not. Roy. Astron. Soc.} {\bf 410}
  (2011) 1993--2002, [\href{http://arxiv.org/abs/1006.0609}{{\tt
  arXiv:1006.0609}}].

\bibitem{Chudaykin:2019ock}
A.~Chudaykin and M.~M. Ivanov, {\it {Measuring neutrino masses with large-scale
  structure: Euclid forecast with controlled theoretical error}},  {\em JCAP}
  {\bf 11} (2019) 034, [\href{http://arxiv.org/abs/1907.06666}{{\tt
  arXiv:1907.06666}}].

\bibitem{Clarkson:2007pz}
C.~Clarkson, B.~Bassett, and T.~H.-C. Lu, {\it {A general test of the
  Copernican Principle}},  {\em Phys. Rev. Lett.} {\bf 101} (2008) 011301,
  [\href{http://arxiv.org/abs/0712.3457}{{\tt arXiv:0712.3457}}].

\bibitem{Hahn_2005}
T.~Hahn, {\it Cuba{\textemdash}a library for multidimensional numerical
  integration},  {\em Computer Physics Communications} {\bf 168} (Jun, 2005)
  78--95.

\bibitem{Philcox:2020vvt}
O.~H.~E. Philcox, M.~M. Ivanov, M.~Simonovi\'c, and M.~Zaldarriaga, {\it
  {Combining Full-Shape and BAO Analyses of Galaxy Power Spectra: A
  1.6\textbackslash{}\% CMB-independent constraint on H$_0$}},  {\em JCAP} {\bf
  05} (2020) 032, [\href{http://arxiv.org/abs/2002.04035}{{\tt
  arXiv:2002.04035}}].

\bibitem{Ivanov:2021kcd}
M.~M. Ivanov, O.~H.~E. Philcox, T.~Nishimichi, M.~Simonovi\'c, M.~Takada, and
  M.~Zaldarriaga, {\it {Precision analysis of the redshift-space galaxy
  bispectrum}},  {\em Phys. Rev. D} {\bf 105} (2022), no.~6 063512,
  [\href{http://arxiv.org/abs/2110.10161}{{\tt arXiv:2110.10161}}].

\bibitem{Philcox:2022frc}
O.~H.~E. Philcox, M.~M. Ivanov, G.~Cabass, M.~Simonovi\'c, M.~Zaldarriaga, and
  T.~Nishimichi, {\it {Cosmology with the redshift-space galaxy bispectrum
  monopole at one-loop order}},  {\em Phys. Rev. D} {\bf 106} (2022), no.~4
  043530, [\href{http://arxiv.org/abs/2206.02800}{{\tt arXiv:2206.02800}}].

\bibitem{DAmico:2020tty}
G.~D'Amico, Y.~Donath, L.~Senatore, and P.~Zhang, {\it {Limits on clustering
  and smooth quintessence from the EFTofLSS}},  {\em JCAP} {\bf 03} (2024) 032,
  [\href{http://arxiv.org/abs/2012.07554}{{\tt arXiv:2012.07554}}].

\bibitem{Ivanov:2020ril}
M.~M. Ivanov, E.~McDonough, J.~C. Hill, M.~Simonovi\'c, M.~W. Toomey,
  S.~Alexander, and M.~Zaldarriaga, {\it {Constraining Early Dark Energy with
  Large-Scale Structure}},  {\em Phys. Rev. D} {\bf 102} (2020), no.~10 103502,
  [\href{http://arxiv.org/abs/2006.11235}{{\tt arXiv:2006.11235}}].

\bibitem{Piga:2022mge}
L.~Piga, M.~Marinucci, G.~D'Amico, M.~Pietroni, F.~Vernizzi, and B.~S. Wright,
  {\it {Constraints on modified gravity from the BOSS galaxy survey}},  {\em
  JCAP} {\bf 04} (2023) 038, [\href{http://arxiv.org/abs/2211.12523}{{\tt
  arXiv:2211.12523}}].

\bibitem{Taule:2024bot}
P.~Taule, M.~Marinucci, G.~Biselli, M.~Pietroni, and F.~Vernizzi, {\it
  {Constraints on dark energy and modified gravity from the BOSS Full-Shape and
  DESI BAO data}},  \href{http://arxiv.org/abs/2409.08971}{{\tt
  arXiv:2409.08971}}.

\bibitem{DAmico_pybird}
G.~D'Amico, L.~Senatore, and P.~Zhang, {\it {Limits on $w$CDM from the EFTofLSS
  with the PyBird code}},  {\em JCAP} {\bf 01} (2021) 006,
  [\href{http://arxiv.org/abs/2003.07956}{{\tt arXiv:2003.07956}}].

\bibitem{Bailoni:2016ezz}
A.~Bailoni, A.~Spurio~Mancini, and L.~Amendola, {\it {Improving Fisher matrix
  forecasts for galaxy surveys: window function, bin cross-correlation, and bin
  redshift uncertainty}},  {\em Mon. Not. Roy. Astron. Soc.} {\bf 470} (2017),
  no.~1 688--705, [\href{http://arxiv.org/abs/1608.00458}{{\tt
  arXiv:1608.00458}}].

\end{thebibliography}\endgroup
 \label{lastpage}
\end{document}